\documentclass[aps,prd,superscriptaddress,showpacs]{revtex4}
\usepackage{graphicx}
\usepackage{epstopdf}
\usepackage{amsmath}
\usepackage{amsfonts}
\usepackage{amssymb}
\usepackage{latexsym}

\begin{document}
\title{$(1+3)$D topological superconductors: screening and confinement in presence of external fields}
\author{Patricio Gaete} \email{patricio.gaete@usm.cl} 
\affiliation{Departamento de F\'{i}sica and Centro Cient\'{i}fico-Tecnol\'ogico de Valpara\'{i}so, Universidad T\'{e}cnica Federico Santa Mar\'{i}a, Valpara\'{i}so, Chile}
\author{Jos\'{e} A. Helay\"{e}l-Neto}\email{helayel@cbpf.br}
\affiliation{Centro Brasileiro de Pesquisas F\'{i}sicas (CBPF), Rio de Janeiro, RJ, Brasil} 
\date{\today}

\begin{abstract}
Adopting the gauge-invariant and path-dependent variables formalism, we compute the interaction energy for
a topological field theory describing  $(1+3)$D  topological superconductors in presence of external fields. As a result, in the case of a constant electric field- strength expectation value, we show that the interaction energy describes a purely screening phase, encoded in a Yukawa potential. On the other hand, in the case of a constant magnetic field-strength and for a very small Josephson coupling constant, the particle-antiparticle binding potential displays a linear term leading to the confinement of static charge probes along with a screening contribution.
\end{abstract}
\maketitle

\section{Introduction}

Over the past years, topological states of matter have become the focus of intense theoretical and experimental research activity \cite{Kane,Zhang}. It is worth noticing that these states are characterized by topological properties rather than usual properties such as symmetries preserved (or broken) by some order parameter. Two of these, which are topological insulators and topological superconductors, have probably enjoyed the greatest popularity. Incidentally, it is of interest note here that topological insulators have been experimentally realized in various materials \cite{Kane,Zhang}. 

We also draw attention to the fact that these states are described by a low-energy field theory which is a topological field theory. In fact, we mention that, a couple of years ago, a topological field theory description of $(1+3)$D topological superconductors has been proposed \cite{WittenZhangQi}. This effective description of a topological superconductor (TS) is expressed by means of a topological coupling between the electromagnetic field and the superconducting phase fluctuation, similarly to the coupling of axions with an Abelian gauge field. It should, however, be highlighted here that this topological field theory also contains a Josephson coupling. In this perspective, it may be mentioned that the presence of this coupling has been discussed in \cite{Nogueira}, in order to distinguish a TS from an ordinary superconductor in the framework of a mean-field analysis.
We further notice that, recently, an interesting approach on this issue has been proposed in \cite{Chung}, which includes a Josephson coupling between topological superconductors and s-wave superconductors.

On the other hand, it is also important to recall that the ideas of screening and confinement play an important role in gauge theory. In this connection, as well- known, the interaction energy between static charge probes is an object of utmost importance, and its physical content can be understood whenever a correct separation of the physical degrees of freedom is carried out.

Motivated by these observations, the purpose of the present contribution is to further elaborate on the physical content of this $(1+3)$-dimensional effective action associated to the description of a TS. To be more precise, our aim is to have some additional understanding of the consequences of including the Josephson coupling term on physical observables from a somewhat different perspective. To accomplish our analysis, we shall use the gauge-invariant but path-dependent variables formalism along the lines of \cite{Nonlinear,Logarithmic,Nonlinear2}, which is an alternative to the Wilson loop approach. When we take this road to compute the static potential in the presence of an external field strength, which can be either electric or magnetic, the result of the calculation is rather unexpected in the magnetic case. In fact, when the Josephson constant coupling is very small, it is shown that the interaction energy is the superposition of a Yukawa and a linear potential, that is, a confining regime between static charges comes out. On the other hand, in the case of a constant electric field-strength value, the static potential keeps its Yukawa character.

Our work is organized according to the following outline: in Section II, we recall the salient features of the model under consideration. In Section III, we compute the interaction energy for both magnetic and electric cases. Finally, some Concluding Remarks are cast in Sec. IV. 

In our conventions, the signature of the metric is ($+1,-1,-1,-1$).

\section{Some aspects of topological superconductors in $(1+3)$ D} 

We now turn to the problem of considering the interaction energy between static point-like sources for a TS in $(1+3)$ dimensions. To do this, we shall compute the expectation value of the energy operator $H$ in the physical state, $\left| \Phi  \right\rangle$, describing the sources, which we will denote by ${\left\langle H \right\rangle _\Phi }$. However, before going to the derivation of the interaction potential, we will describe some features of the model under consideration. For this purpose, we restrict our attention to the effective action, which in the case of two Fermi surfaces, reads \cite{WittenZhangQi}:
\begin{equation}
{\cal L} =\frac{{{e^2}\theta }}{{64{\pi ^2}}}{\varepsilon ^{\mu \nu \sigma \tau }}{F_{\mu \nu }}{F_{\sigma \tau }} - \frac{1}{4}{F_{\mu \nu }}{F^{\mu \nu }} + \frac{1}{2}{\rho _L}{\left( {{\partial _\mu }{\theta _L} - 2e{A_\mu }} \right)^2} + \frac{1}{2}{\rho _R}{\left( {{\partial _\mu }{\theta _R} - 2e{A_\mu }} \right)^2} + J\cos \theta, \label{Top05}
\end{equation}
where the charge, $2e$, is the charge of the condensate. The first term (topological) contains a scalar field, $\theta  = {\theta _L} - {\theta _R}$, where ${\theta _L}$ and ${\theta _R}$ are functions on two Fermi surfaces (left and right). While  ${\rho _L}$ and ${\rho _R}$ refer to the density of Cooper pairs on the two Fermi surfaces. The last term describes a Josephson coupling.

It may be noticed that, by reshuffling below, 
\begin{equation}
{B_\mu } = {A_\mu } - \frac{1}{{2e}}\frac{{\left( {{\rho _L}{\partial _\mu }{\theta _L} + {\rho _R}{\partial _\mu }{\theta _R}} \right)}}{{\left( {{\rho _L} + {\rho _R}} \right)}}, \label{Top10}
\end{equation}
equation (\ref{Top05}) can alternatively be rewritten in the form 
\begin{equation}
{\cal L} = \frac{{{e^2}\theta }}{{64{\pi ^2}}}{\varepsilon ^{\mu \nu \sigma \tau }}{F_{\mu \nu }}{F_{\sigma \tau }} - \frac{1}{4}{F_{\mu \nu }}{F^{\mu \nu }} + 2{e^2}\left( {{\rho _L} + {\rho _R}} \right){A_\mu }{A^\mu } + \frac{1}{2}\frac{{{\rho _L}{\rho _R}}}{{\left( {{\rho _L} + {\rho _R}} \right)}}({\partial _\mu }\theta )\left( {{\partial ^\mu }\theta } \right) + J\cos \theta. \label{Top15}
\end{equation}

Before going ahead, it is very convenient to restore the gauge invariance of equation (\ref{Top15}), for it is the gauge invariance that generally ensures unitarity and renormalizability in most quantum field-theoretical models. To do this, we shall adopt our earlier procedure \cite{Nonlinear2}. In this way, we easily verify that the canonical momenta read ${\Pi ^\mu } =  - {F^{0\mu }} + \frac{{{e^2}\theta }}{{8{\pi ^2}}}{\tilde F^{0\mu }}$, which results in the usual primary constraint ${\Pi ^0 }=0$, and ${\Pi ^i} = {E^i} - \frac{{{e^2}\theta }}{{8{\pi ^2}}}{B^i}$, where ${\tilde F^{\mu \nu }} = \frac{1}{2}{\varepsilon _{\mu \nu \lambda \rho }}{F^{\lambda \rho }}$. Then, the canonical Hamiltonian for the electromagnetic sector becomes
\begin{equation}
{H_C} = \int {{d^3}x} \left\{ { - {A_0}\left( {{\partial _i}{\Pi ^i} + {m^2}{A^0}} \right) + \frac{1}{2}{{\bf \Pi} ^2} + \frac{1}{2}{{\bf B}^2} + \frac{1}{2}{{\left( {\frac{{{e^2}\theta }}{{8{\pi ^2}}}} \right)}^2}{{\bf B}^2} + {m^2}{{\bf A}^2} + \frac{{{e^2}\theta }}{{8{\pi ^2}}}{\bf \Pi} \cdot {\bf B}} \right\}.  \label{Top20}
\end{equation}
Requiring the primary constraint $\Pi^{0}$ to be preserved in time, one obtains the constraint $\Gamma  \equiv {\partial _i}{\Pi ^i} + {m^2}{A^0} = 0$. Evidently, both constraints are second class. As already expressed, to convert the second-class system into a first-class one, we shall adopt the procedure described in \cite{Nonlinear2}. To this end, we enlarge the original phase space by introducing a canonical pair of fields $\xi$ and ${\Pi _\xi }$. 
It follows, therefore, that a new set of first class constraints can be defined in this extended space: ${\Lambda _1} \equiv {\Pi _0} + {m^2}\xi =0$ and ${\Lambda _2} \equiv \Gamma  + {\Pi _\xi } = 0$. As a consequence, the new constraints are first-class and, consequently, the gauge symmetry of the theory under consideration is restored.
From this, the new effective Lagrangian density, after integrating out the $\xi$-fields, takes the form
\begin{equation}
{\cal L} =  - \frac{1}{4}{F_{\mu \nu }}\left( {1 + \frac{{{m^2}}}{\Delta }} \right){F^{\mu \nu }} + \frac{{{e^2}\theta }}{{32{\pi ^2}}}{F_{\mu \nu }}{\tilde F^{\mu \nu }} + \frac{1}{2}\frac{{{\rho _L}{\rho _R}}}{{\left( {{\rho _L} + {\rho _R}} \right)}}{\partial _\mu }\theta {\partial ^\mu }\theta  + J\cos \theta, \label{Top25}
\end{equation}
where $\Delta  = {\partial _\mu }{\partial ^\mu }$ and ${m^2} = 2{e^2}\left( {{\rho _L} + {\rho _R}} \right)$.

In order to handle the cosine term in (\ref{Top25}), we shall now consider the expansion near $\theta=\pi$ \cite{Nogueira}. In such a case, after integrating out the $\theta$-fields, we get an effective action for the ${A_\mu }$-field, which reads as follows:
\begin{equation}
{\cal L} =  - \frac{1}{4}{F_{\mu \nu }}\left( {1 + \frac{{{m^2}}}{\Delta }} \right){F^{\mu \nu }} + \frac{1}{2}\left( {J\pi  + \lambda {F_{\mu \nu }}{\tilde F^{\mu \nu }}} \right)\frac{1}{{\left( {{\chi ^2}\Delta  + J} \right)}}\left( {J\pi  + \lambda  {F_{\sigma \rho }}{\tilde F^{\sigma \rho }}} \right),\label{Top30}
\end{equation}
where ${\chi ^2} = \frac{{{\rho _L}{\rho _R}}}{{\left( {{\rho _L} + {\rho _R}} \right)}}$ and $\lambda  = \frac{{{e^2}}}{{32{\pi ^2}}}$.\\

This new effective description provides us with a suitable starting point to study the energy interaction between static charge probes.

\section{Interaction energy}

Having established the new effective Lagrangian, we can compute the interaction energy between static point-like sources for the model under consideration. It should, however, be noted that the system described by the Lagrangian 
(\ref{Top30}) is a system with nonlocal time derivatives. Nevertheless, we stress that this paper is aimed at studying the static potential of the theory (\ref{Top30}), so that $\Delta$ can be replaced by $- {\nabla ^2}$.

Next, in order to study quantum properties of the electromagnetic field in the presence of external electromagnetic fields, we split the ${A_\mu }$-field as the sum of a classical background, $\left\langle {{A_\mu }} \right\rangle$, and a small quantum fluctuation, ${a_\mu }$, that is: ${A_\mu } = \left\langle {{A_\mu }} \right\rangle  + {a_\mu }$. In this manner, up to quadratic terms in the fluctuations, the previous Lagrangian density can be brought to the form
\begin{equation}
{\cal L} =  - \frac{1}{4}{f_{\mu \nu }}\left( {1 - \frac{{{m^2}}}{{{\nabla ^2}}}} \right){f^{\mu \nu }} - \frac{{{\lambda ^2}}}{2}{v^{\mu \nu }}{f_{\mu \nu }}\frac{1}{{\left( {{\chi ^2}{\nabla ^2} - J} \right)}}{v^{\lambda \rho }}{f_{\lambda \rho }}, \label{Top35}
\end{equation}
where ${f_{\mu \nu }} = {\partial _\mu }{a_\nu } - {\partial _\nu }{a_\mu }$ and ${\varepsilon ^{\mu \nu \alpha \beta }}\left\langle {{F_{\mu \nu }}} \right\rangle  \equiv {v^{\alpha \beta }}$, in this case ${v^{\alpha \beta }}$ satisfies ${\varepsilon ^{\mu \nu \alpha \beta }}{v_{\mu \nu }}{v_{\alpha \beta }} = 0$.

\subsection{Magnetic case}

With the foregoing information, we proceed to compute the interaction energy in the ${v^{0i}} \ne 0$ and ${v^{ij}} = 0$ case (referred to as the magnetic one in what follows), via the expectation value of the energy operator $H$ in the physical state $\left| \Phi  \right\rangle$. 

We begin our discussion by obtaining the Hamiltonian. We first observe that the canonical momenta read $\Pi ^0  = 0$ and ${\Pi _i} = \left( {1 - \frac{{{m^2}}}{{{\nabla ^2}}}} \right){D_{ij}}{E_j}$, where ${E_i} = {f_{i0}}$ and ${D_{ij}} = \left( {{\delta _{ij}} - \frac{{{\lambda ^2}}}{{{\chi ^2}}}{v_{i0}}\frac{{{\nabla ^2}}}{{\left( {{\nabla ^2} - {m^2}} \right)\left( {{\nabla ^2} - {\mu ^2}} \right)}}{v_{j0}}} \right)$. This allows us to write the following canonical Hamiltonian 
\begin{equation}
{H_C} = \int {{d^3}x} \left\{ { - {a_0}{\partial ^i}{\Pi _i} + \frac{1}{2}{\Pi ^i}\left[ {\frac{{\left( {{\nabla ^2} - {\mu ^2}} \right)}}{{\left\{ {\left[ {{\nabla ^2} - \left( {{m^2} + {\mu ^2} + \frac{{{\lambda ^2}}}{{{\chi ^2}}}{{\bf v}^2}} \right)} \right] + \frac{{{m^2}{\mu ^2}}}{{{\nabla ^2}}}} \right\}}}} \right]{\Pi ^i} + \frac{1}{2}{B^i}\left( {1 - \frac{{{m^2}}}{{{\nabla ^2}}}} \right){B^i}} \right\},  \label{Top40}
\end{equation}
where ${{\bf v}^2} = {v^{i0}}{v^{i0}}$ and ${\mu ^2} = \frac{J}{{{\lambda ^2}}}$.

Demanding that the primary constraint $\Pi^{0}=0$ be preserved in the course of time, one obtains the secondary Gauss law constraint $\Gamma _1 \equiv \partial _i \Pi ^i  = 0$. The corresponding extended Hamiltonian that generates translations in time then reads $  
H = H_C  + \int {d^3 x} \left( {u_0(x) \Pi_0(x)  + u_1(x) \Gamma _1(x) } \right)$, where $u_o(x)$ and $u_1(x)$ are the Lagrange multiplier utilized to implement the constraints. Moreover, it follows from this Hamiltonian that $\dot A_0 \left( x \right) = \left[ {A_0 \left( x \right),H} \right] = u_0 \left( x \right)$, which is completely arbitrary function. Since ${\Pi ^0} = 0$ always, we discard both $A_0$ and $\Pi_0$ from the theory. Thus the Hamiltonian takes the form
\begin{equation}
H = \int {{d^3}x} \left\{ {  w(x){\partial ^i}{\Pi _i} + \frac{1}{2}{\Pi ^i}\left[ {\frac{{\left( {{\nabla ^2} - {\mu ^2}} \right)}}{{\left\{ {\left[ {{\nabla ^2} - \left( {{m^2} + {\mu ^2} + \frac{{{\lambda ^2}}}{{{\chi ^2}}}{{\bf v}^2}} \right)} \right] + \frac{{{m^2}{\mu ^2}}}{{{\nabla ^2}}}} \right\}}}} \right]{\Pi ^i} + \frac{1}{2}{B^i}\left( {1 - \frac{{{m^2}}}{{{\nabla ^2}}}} \right){B^i}} \right\},  \label{Top45}
\end{equation}
where $w(x) = u_1 (x) - A_0 (x)$.

In order to quantize the theory we introduce a gauge condition on the vector potential such that the full set of constraints becomes second class. A particularly convenient gauge-fixing condition is
\begin{equation}
\Gamma _2 \left( x \right) \equiv \int\limits_{C_{\xi x} } {dz^\nu
} A_\nu \left( z \right) \equiv \int\limits_0^1 {d\lambda x^i }
A_i \left( {\lambda x} \right) = 0, \label{Top50}
\end{equation}
where  $\lambda$ $(0\leq \lambda\leq1)$ is the parameter describing
the spacelike straight path $ z^i = \xi ^i  + \lambda \left( {x -\xi } \right)^i $, and $ \xi $ is a fixed point (reference point). There is no essential loss of generality if we restrict our considerations to $ \xi ^i=0 $. Hence the only nontrivial Dirac bracket for the canonical variables is given by
\begin{equation}
\left\{ {A_i \left( x \right),\Pi ^j \left( y \right)} \right\}^ *
= \delta _i^j \delta ^{\left( 3 \right)} \left( {x - y} \right) -
\partial _i^x \int\limits_0^1 {d\lambda x^i } \delta ^{\left( 3
\right)} \left( {\lambda x - y} \right). \label{Top55}
\end{equation}

We now go over into the calculation of the interaction energy. As mentioned above, to complete this task, we shall compute the expectation value of the energy operator $H$ in the physical state $\left| \Phi  \right\rangle$. Following Dirac \cite{Dirac}, we write the physical state $\left| \Phi  \right\rangle$ as
\begin{equation}
\left| \Phi  \right\rangle  \equiv \left| {\bar \Psi ({\bf y})\Psi ({{\bf y}^ \prime })} \right\rangle  = \bar \psi ({\bf y})\exp (ie\int_{{{\bf y}^ \prime }}^{\bf y} {d{z^i}{A_i}(z)} )\psi ({{\bf y}^ \prime })\left| 0 \right\rangle,                              
\label{Top60}
\end{equation}
where $\left| 0 \right\rangle$ is the physical vacuum state and the
line integral appearing in the above expression is along a spacelike
path starting at ${\bf y}\prime$ and ending at $\bf y$, on a fixed
time slice. The above expression clearly shows that, each of the states $(\left| \Phi  \right\rangle)$, 
represents a fermion-antifermion pair surrounded by a cloud of gauge fields to maintain gauge invariance.

Taking into account the above Hamiltonian structure, we observe that
\begin{equation}
\Pi _i \left( x \right)\left| {\overline \Psi \left( \mathbf{y }\right)\Psi
\left( {\mathbf{y}^ \prime } \right)} \right\rangle = \overline \Psi \left( 
\mathbf{y }\right)\Psi \left( {\mathbf{y}^ \prime } \right)\Pi _i \left( x
\right)\left| 0 \right\rangle + e\int_ {\mathbf{y}}^{\mathbf{y}^ \prime } {\
dz_i \delta ^{\left( 3 \right)} \left( \mathbf{z - x} \right)} \left| \Phi
\right\rangle.  \label{Top70}
\end{equation}
Therefore, the interaction energy can be written as
\begin{equation}
V\equiv{\left\langle H \right\rangle _\Phi } = {\left\langle H \right\rangle _0} + {\left\langle H \right\rangle _1} + {\left\langle H \right\rangle _2}, \label{Top75}
\end{equation}
where $\left\langle H \right\rangle _0  = \left\langle 0 \right|H\left| 0 \right\rangle$. The $\left\langle H \right\rangle _1$, $\left\langle H \right\rangle _2$ terms are given by
\begin{equation}
{\left\langle H \right\rangle _1} = \frac{1}{2}\left\langle \Phi  \right|\int {{d^3}x} {\Pi ^i}\frac{{{\nabla ^4}}}{{\left[ {{\nabla ^4} - \left( {{m^2} + {\mu ^2} + \frac{{{\lambda ^2}}}{{{\chi ^2}}}{{\bf v}^2}} \right){\nabla ^2} + {m^2}{\mu ^2}} \right]}}{\Pi ^i}\left| \Phi  \right\rangle, \label{Top80} 
\end{equation}
and
\begin{equation}
{\left\langle H \right\rangle _2} =  - \frac{{{\mu ^2}}}{2}\left\langle \Phi  \right|\int {{d^3}x} {\Pi ^i}\frac{{{\nabla ^2}}}{{\left[ {{\nabla ^4} - \left( {{m^2} + {\mu ^2} + \frac{{{\lambda ^2}}}{{{\chi ^2}}}{{\bf v}^2}} \right){\nabla ^2} + {m^2}{\mu ^2}} \right]}}{\Pi ^i}\left| \Phi  \right\rangle. \label{Top85} 
\end{equation}
Using equation (\ref{Top70}), we see that the potential for two opposite charges, localized at ${\bf y}$ and ${\bf {y^\prime}}$, takes the form   
\begin{equation}
V = - \frac{{{e^2}}}{{4\pi }}\frac{{\left( {M_1^2 - {\mu ^2}} \right)}}{{\left( {M_1^2 - M_2^2} \right)}}\frac{{{e^{ - {M_1}L}}}}{L} - \frac{{{e^2}}}{{4\pi }}\frac{{\left( {{\mu ^2} - M_2^2} \right)}}{{\left( {M_1^2 - M_2^2} \right)}}\frac{{{e^{ - {M_2}L}}}}{L}, \label{Top90}
\end{equation}
where $|{\bf y} - {{\bf y}^ \prime }| = L$, while $M_1^2$ and $M_2^2$ are given by $M_1^2 = \frac{1}{2}\left[ {\left( {{m^2} + {\mu ^2} + \frac{{{\lambda ^2}}}{{{\chi ^2}}}{{\bf v}^2}} \right) + \sqrt {{{\left( {{m^2} + {\mu ^2} + \frac{{{\lambda ^2}}}{{{\chi ^2}}}{{\bf v}^2}} \right)}^2} - 4{m^2}{\mu ^2}} } \right]$ and $M_2^2 = \frac{1}{2}\left[ {\left( {{m^2} + {\mu ^2} + \frac{{{\lambda ^2}}}{{{\chi ^2}}}{{\bf v}^2}} \right) - \sqrt {{{\left( {{m^2} + {\mu ^2} + \frac{{{\lambda ^2}}}{{{\chi ^2}}}{{\bf v}^2}} \right)}^2} - 4{m^2}{\mu ^2}} } \right]$.

In so doing, we see that this effective theory describes exactly a screening phase, encoded in the Yukawa-type potentials.

On the other hand, it is of interest to note that in the case of a very small $\mu$, from equation $(\ref{Top45})$ it now follows that the interaction energy can be written in the form
\begin{equation}
V\equiv{\left\langle H \right\rangle _\Phi } = {\left\langle H \right\rangle _0} + {\left\langle H \right\rangle _1} + {\left\langle H \right\rangle _2} + {\left\langle H \right\rangle _3}, \label{Top90}
\end{equation}
where
\begin{equation}
{\left\langle H \right\rangle _1} = \frac{1}{2}\left\langle \Phi  \right|\int {{d^3}x} {\Pi ^i}\frac{{{\nabla ^2}}}{{\left( {{\nabla ^2} - {M^2}} \right)}}{\Pi ^i}\left| \Phi  \right\rangle, \label{Top95}  
\end{equation}
\begin{equation}
{\left\langle H \right\rangle _2} =  - {\mu ^2}\left( {1 - \frac{{{m^2}}}{{{M^2}}}} \right)\left\langle \Phi  \right|\int {{d^3}x} {\Pi ^i}\frac{1}{{\left( {{\nabla ^2} - {M^2}} \right)}}{\Pi ^i}\left| \Phi  \right\rangle, \label{Top100}
\end{equation}
\begin{equation}
{\left\langle H \right\rangle _3} = \frac{1}{2}\frac{{{m^2}{\mu ^2}}}{{{M^2}}}\left\langle \Phi  \right|\int {{d^3}x} {\Pi ^i}\frac{{{\nabla ^2}}}{{{{\left( {{\nabla ^2} - {M^2}} \right)}^2}}}{\Pi ^i}\left| \Phi  \right\rangle, \label{Top105}
\end{equation}
where ${M^2} = {m^2} + {\mu ^2} + \frac{{{\lambda ^2}}}{{{\chi ^2}}}{{\bf v}^2}$.

Once again, following our earlier procedure, we see that the potential for two opposite charges located at ${\bf y}$ and ${\bf y}^{\prime}$ takes the form
\begin{equation}
V =  - \frac{{{e^2}}}{{4\pi }}\frac{{{e^{ - ML}}}}{L} + \frac{{{e^2}}}{{4\pi }}{\mu ^2}\left( {1 - \frac{{{m^2}}}{{{M^2}}}} \right)\ln \left( {1 + \frac{{{\Lambda ^2}}}{{{M^2}}}} \right)L - \frac{{{e^2}}}{{8\pi }}\frac{{{m^2}{\mu ^2}}}{{{M^3}}}{e^{ - ML}}, \label{Top110}
\end{equation}
where $\Lambda$ is a cutoff parameter in momentum space. The effective photonic model cast in eq. $(\ref{Top30})$, from which we have started off to compute the particle-antiparticle potential $(\ref{Top110})$  , takes already into account the integration over the fermionic sector. So, the process of fermion pair condensation is implicitly included in the potential. Therefore, the cutoff $\Lambda$ must be of the order of the pair condensation scale, that is,
the BCS-gap: $\Lambda  \sim$ EBCS. Above the BCS scale, our results cannot be applied. By considering the typical values of the pair binding energies, $\Lambda$ is of the order of eV. It is worthwhile tohighlight that we observethat a very small Josephson coupling constant induces a screening part, encoded in the Yukawa potential, plus a linear confining potential.

\subsection{Electric case}

We now wish to extend what we have worked out to the case ${v^{0i}}=0$ and ${v^{ij}} \ne 0$ (referred to as the electric case one in what follows). In such a case, the Lagrangian density reads
\begin{equation}
{\cal L} =  - \frac{1}{4}{f_{\mu \nu }}\left( {1 - \frac{{{m^2}}}{{{\nabla ^2}}}} \right){f^{\mu \nu }} - \frac{{{\lambda ^2}}}{{2{\chi ^2}}}{v^{ij}}{f_{ij}}\frac{1}{{\left( {{\nabla ^2} - {\mu ^2}} \right)}}{v^{kl}}{f_{kl}}.  \label{Top115}
\end{equation}

The Lagrangian density above will be the starting point of the Dirac constrained analysis. The canonical momenta following from equation (\ref{Top115}) are ${\Pi ^\mu } = \left( {1 - \frac{{{m^2}}}{{{\nabla ^2}}}} \right){f^{\mu 0}}$, which results in the usual primary constraint ${\Pi ^0} = 0$ and
${\Pi ^i} = (1 - \frac{{{m^2}}}{{{\nabla ^2}}}){f^{i0}}$. The canonical Hamiltonian can be worked as usual and is given by
\begin{equation}
{H_C} = \int {{d^3}x} \left\{ {{\Pi ^i}{\partial _i}{A_0} + \frac{1}{2}{\Pi ^i}\frac{{{\nabla ^2}}}{{\left( {{\nabla ^2} - {m^2}} \right)}}{\Pi ^i} + \frac{1}{2}{B^i}\left( {1 - \frac{{{m^2}}}{{{\nabla ^2}}}} \right){B^i} + \frac{{{\lambda ^2}}}{{2{\chi ^2}}}{\varepsilon _{ijm}}{\varepsilon _{k\ln }}{v^{ij}}{B^m}\frac{1}{{\left( {{\nabla ^2} - {\mu ^2}} \right)}}{v^{kl}}{B^n}} \right\}. \label{Top120}
\end{equation}

Time conservation of the primary constraint leads to the secondary constraint ${\Gamma _1}(x) \equiv {\partial _i}{\Pi ^i} = 0$, and the time stability of the secondary constraint does not induce more constraints.

By proceeding in the same way as in the previous subsection, we obtain the static potential for two opposite charges located at ${\bf y}$ and ${\bf y}^\prime$:
\begin{equation}
V =  - \frac{{{e^2}}}{{4\pi }}\frac{{{e^{ - mL}}}}{L}. \label{Top125}
\end{equation}

We immediately see that the confining potential between static charges vanishes in this case. In other words: in this case, the model exactly describes a screening phase.

\section{Concluding Remarks}

In summary, we have considered the recently-proposed topological field theory \cite{WittenZhangQi}, which describes  $(1+3)$D  topological superconductors from a different perspective. We have followed here the gauge-invariant and path-dependent variables formalism in the presence of external fields. Once again, a correct identification of physical degrees of freedom has been fundamental for understanding the physics hidden in gauge theories. It has been shown that, in the case of a constant electric field-strength expectation value, the interaction energy describes a purely screening phase, encoded in a Yukawa potential. Interestingly enough, in the case of a constant magnetic field-strength and for a very small Josephson coupling constant, the interaction energy profile contains a linear term leading to the confinement of static charge probes along with a screening contribution.

The model we investigate here is supported by a five-dimensional scenario and its corresponding $(1+3)$-dimensional holographic projection \cite{WittenZhangQi}. In this framework, we would like to conclude our paper by raising an issue that we are now working on \cite{GHO}: the possibility that fermion condensation takes place already in $5$ dimensions, before the 4-dimensional holographic projection is taken. Taking this viewpoint opens up a non-trivial discussion in connection with the fermion mass term in $(1+4)$D: actually, a Dirac mass term in $5$ dimensions explicitly breaks parity symmetry. To come over this problem, we have to double the spinor representation associated to the electron field. In so doing, P-invariance can suitably be implemented through a mixed mass term involving the two four-component spinors assembled together to describe a (parity-preserving) massive electron in $5$ dimensions. However, in this picture, the mirror fermion present in $5$ dimensions may yield vector condensates that show up in $4$ dimensions as (background) anisotropies and induce non-trivial effects in the superconducting phase of the material. We are interested in this particular point, once we assume that the truly fundamental physics underneath a TS takes place in $5$ dimensions and, then, to work in this landscape, we have to face the problem of the fermion mass pointed out above \cite{GHO} and the anisotropies which may be induced whenever we lower the space-time dimension.

\section{ACKNOWLEDGMENTS}
P. G. was partially supported by Fondecyt (Chile) grant 1130426 and by Proyecto Basal FB0821.


\begin{thebibliography}{}

\bibitem{Kane}  M.~Z.~Hasan and C.~L.~Kane, Rev.\ Mod.\ Phys.\  {\bf 82}, 3045 (2010).

\bibitem{Zhang} X.~L.~Qi and S.~C.~Zhang, Rev.\ Mod.\ Phys.\  {\bf 83}, 1057 (2011).
 
\bibitem{WittenZhangQi} X.~L.~Qi, E.~Witten and S.~C.~Zhang, Phys.\ Rev.\ B {\bf 87}, 134519 (2013).
  
\bibitem{Nogueira} F.~S.~Nogueira, A.~Sudb$\phi$  and I.~Eremin, Phys.\ Rev.\ B {\bf 92}, 224507 (2015).

\bibitem{Chung} S.~B.~Chung, J.~Horowitz and X.~L.~Qi, Phys.\ Rev.\ B {\bf 88}, 214514 (2013).

\bibitem{Nonlinear} P.~Gaete and J.~Helay\"el-Neto, Eur.\ Phys.\ J.\ C {\bf 74}, 3182 (2014).

\bibitem{Logarithmic}  P.~Gaete and J.~Helay\"el-Neto, Eur.\ Phys.\ J.\ C {\bf 74}, 2816 (2014).

\bibitem{Nonlinear2} P.~Gaete, Adv.\ High Energy Phys.\  {\bf 2016}, 2463203 (2016).

\bibitem{Dirac} P. A. M. Dirac, Can. J. Phys. {\bf 33}, 650 (1955).

\bibitem{GHO}  P.~Gaete, J.~Helay\"el-Neto and L. P. R. Ospedal, work in development.
\end{thebibliography}
\end{document}